\documentclass[11pt,a4paper]{article}
\pdfoutput=1
\usepackage{jcappub}
\usepackage[normalem]{ulem}
\newcommand{\new}[1]{{{#1}}}

\title{Asymmetric velocity anisotropies in remnants of collisionless mergers}
\author[]{Martin Sparre}
\author[]{and Steen H. Hansen}
\affiliation[]{Dark Cosmology Centre, Niels Bohr Institute,\\
University of Copenhagen, Juliane Maries Vej 30, 2100 Copenhagen, Denmark}

\emailAdd{sparre@dark-cosmology.dk}
\emailAdd{hansen@dark-cosmology.dk}

\abstract{
Dark matter haloes in cosmological N-body simulations are affected by processes such as mergers, accretion and the gravitational interaction with baryonic matter. \new{Typically the analysis of dark matter haloes is performed in spherical or elliptical bins and the velocity distributions are often assumed to be constant within those bins}. 
\new{However, the velocity anisotropy, which describes differences between the radial and tangential velocity dispersion, has recently been show to have a strong dependence on direction in the triaxial halos formed in cosmological simulations.} \new{In this study} we derive properties of particles in cones parallel or perpendicular to the collision axis of merger remnants. We find that the velocity anisotropy has a strong dependence on direction. The finding that the direction-dependence of the velocity anisotropy of a halo depends on the merger history, \new{explains the existence of such trends in cosmological simulations. It also explains why a large diversity is seen in the velocity anisotropy profiles in the outer parts of high-resolution simulations of cosmological haloes.}
}


\begin{document}
\maketitle

\section{Introduction}

In the $\Lambda$CDM-paradigm the first bound dark matter haloes are formed through the collapse of overdense regions, which have decoupled from the Hubble expansion. These haloes grow when they accrete matter from their surroundings and when they merge. Understanding merger remnants is therefore essential in order to understand the dark matter distribution in the universe.

The dark matter (DM), making up most of the matter in the universe \cite{1998AJ....116.1009R,1999ApJ...517..565P,2011ApJS..192...18K}, consists of particles interacting so weakly, that haloes can be well modelled with collisionless mechanics. An example of evidence for this collisionless nature of DM comes from observations and simulations of the Bullet Cluster \cite{2006ApJ...648L.109C,2006ApJ...652..937B,2007MNRAS.380..911S,2008MNRAS.389..967M}, which consists of two merging galaxy clusters with the collisional baryonic matter separated from the collisionless dark matter.

A difference between collisional and collisionless particles is that collisionless particles tend to have anisotropic velocity distributions, typically parametrised by the velocity anisotropy parameter \cite{2008gady.book.....B},
\begin{align}
\beta (r) \equiv 1-\frac{\sigma_\text{tan}^2 (r)}{2\sigma_\text{rad}^2(r)},
\end{align}
where $\sigma_\text{tan}$ is the total tangential velocity dispersion at a given radius ($r$), and $\sigma_\text{rad}$ is the radial velocity dispersion. The velocity distributions are radially dominated for $\beta > 0$, tangentially dominated for $\beta<0$, and isotropic for $\beta=0$. In cosmological simulations the $\beta$-profiles increase from $\beta\simeq 0$ in the central regions to $\beta\simeq 0.25$ at $r_{-2}$ \new{\citep{2004MNRAS.352..535D,2011MNRAS.tmp..937L}}, which is the radius with a density slope of $\gamma=-2$, where
\begin{align}
\gamma \equiv \frac{d\log \rho}{d\log r}.
\end{align}
At larger radii the $\beta$-profiles vary from halo to halo.

Dark matter haloes from cosmological simulations have universal pseudo-phase-space density profiles following power laws \citep{2001ApJ...563..483T},
\begin{align}
\rho/\sigma^3_\text{rad}(r)\propto r^{-\alpha} \label{psd},
\end{align}
with $\alpha \simeq 1.91$ \citep{2011MNRAS.tmp..937L}. The density profiles are also universal \citep{1996ApJ...462..563N,1997ApJ...490..493N}, with $\gamma\simeq -1$ in the inner parts and $\gamma\simeq -3$ in the outer parts. Typically haloes are parametrised with the NFW-profile, or the Einasto profile \citep{2008MNRAS.387..536G,2010MNRAS.402...21N}. The universality of the density- and the pseudo-phase-space-density-profile in cosmological haloes determines $\sigma_\text{rad}^2 (r)$. The velocity anisotropy, however, remains undetermined.

In an attempt to understand the physics of the velocity anisotropy, several studies have examined $\beta$-profiles in non-cosmological simulations. A linear relation \cite{2006NewA...11..333H,2006JCAP...05..014H} between $\beta$ and $\gamma$ is consistent with the inner parts of cosmological haloes. More recently a 1-dimensional relation (an \emph{attractor}) between $\beta$, $\gamma$ and the slope of the radial velocity dispersion profile has been found in controlled non-cosmological simulations \cite{2010ApJ...718L..68H,2012arXiv1204.2764B}. A tight relation between $\beta$ and $\gamma$, which is predicted by this attractor, is, however, not consistent with the large scatter seen in the outer parts of cosmological haloes. \new{In the two proposed relations, which determines $\beta(r)$ from a given density profile, $\gamma(r)$, it is implicitly assumed that $\beta$ is constant in spherical or elliptical bins.}

\new{In a detailed study of a cosmological halo \citep{2009MNRAS.394..641Z}, it has, however, been found that $\beta$ behaves differently along different axes. The aim of this paper is to show that such direction-dependent $\beta$-profiles can arise because of mergers.}
\new{We will address this issue by studying the velocity anisotropy in different directions of merger remnants, to see whether $\beta$-profiles are direction dependent. In our analysis we will cut out cones of particles in different directions, and then calculate $\beta$ and other variables for the particles inside each cone.} In Section~\ref{MergerSim} we will present merger simulations, and in Section~\ref{ConesSec} we will map the velocity anisotropy parallel and perpendicular to the collision axis and determine whether they are different. We also show how structures are affected by smooth accretion (Section~\ref{Accretion}), and \emph{skymaps} of the $\beta$-profiles are presented (Section~\ref{skymap}). Section~\ref{b-g-relations} discusses relations between $\beta$ and $\gamma$, and Section~\ref{SectionPPSD} examines pseudo-phase-space density profiles. We also discuss the shape of the merger remnants (Section~\ref{shape}).

Throughout this paper we will use units with $G=1$ and let $\log x$ denote the logarithm with base 10.

\section{The merger simulations}\label{MergerSim}

\subsection{Initial conditions}

We are interested in doing a simulation of the merging of two identical dark matter haloes. To do so Eddington's formula \cite{1916MNRAS..76..572E} (with $\beta=0$) is used to generate a halo with a Hernquist density profile \cite{1990ApJ...356..359H}, 
\begin{align}
\rho (r) = \frac{\rho_0}{r/r_\text{s}} \frac{1}{(1+r/r_\text{s})^3}.
\end{align}
We choose $\rho_0=1/2\pi$ and $r_\text{s}=1$, which gives a structure with a total mass of 1. The density profile is truncated at a radius of $10r_\text{s}$. This density profile is chosen because of its simple analytical properties, and because it resembles \cite{2005MNRAS.361..776S} the NFW-profile, which describes haloes in cosmological simulations.

As initial conditions two such structures are generated, and placed $20r_\text{s}$ away from each other along the $x$-axis. $10^6$ collisionless particles are used to represent each halo. The haloes approached each other with an initial velocity of $80\%$ the escape velocity at this distance for an isolated Hernquist halo. Simulations were run with no impact parameter, and with an impact parameter of $10r_\text{s}$ (in the $y$-direction) and with the same initial velocity and initial offset along the $x$-axis.

Simulations were also made with an Osipkov-Merritt model \cite{1979SvAL....5...42O,1985AJ.....90.1027M} (still following the Hernquist profile described above) with the velocity anisotropy given by 
\begin{align}
\beta (r) = \frac{r^2}{r^2+r_\text{a}^2},
\end{align}
where $r_\text{a}$ is the \emph{anisotropy radius}, which we fix to $1.2r_\text{s}$. An anisotropy radius slightly larger than $r_\text{s}$ is chosen to avoid instabilities \cite{1997ApJ...490..136M}.

Simulations with 1:10 mergers were also performed (only for the model with $\beta=0$). A structure, identical to the $\beta=0$ model defined above, was collided with a Hernquist structure with a 10 times lower total mass and a $r_\text{s}$-value and cutoff radius 3 times smaller.

\subsection{Simulation details}
The public version of the N-body simulation code Gadget-2 \cite{2005MNRAS.364.1105S} was used to run the simulations. All particles were collisionless and a spline softening of 0.015 was used in all simulations. All simulations were run for 300 time units.

\section{Analysing particles in cones}\label{ConesSec}

\begin{figure}
\centering
\includegraphics[width=0.8\textwidth]{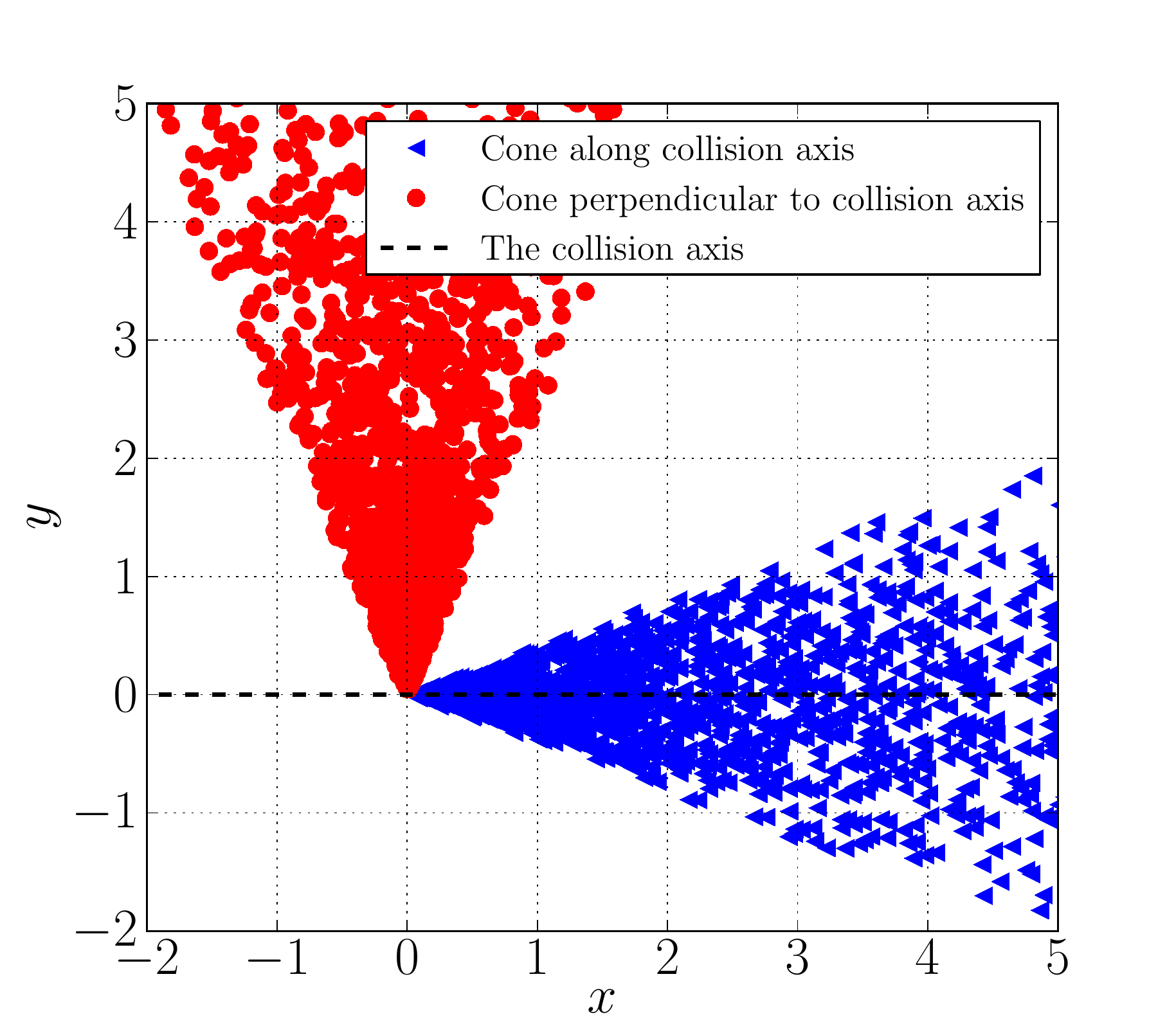}
\caption{Properties of particles inside a cone parallel and a cone perpendicular to the collision axis are studied. Particles outside the cones are not shown.}
\label{Cones}
\end{figure}

To map the velocity anisotropy in the remnants, we defined cones pointing along the $x$-, $y$- and $z$-axis. The angle between the position vector ($\mathbf{r}$) of a particle and a unit vector ($\hat{\mathbf{n}}$) pointing in the direction of a cone is $\theta = \arccos (\hat{\mathbf{n}}\cdot \mathbf{r} / |\mathbf{r}|)$. We chose the cones to have an apex angle of $45^\circ$, so a particle is inside a cone if $\theta \le 22.5^\circ$. The cones are centered in the center of the merger remnants, i.e. at the position of the particle with the lowest potential. Figure~\ref{Cones} shows the particles in two cones for one of the merger remnants.

\subsection{Major mergers}\label{major_mergers}

\begin{figure}
\centering
\includegraphics[width=\textwidth]{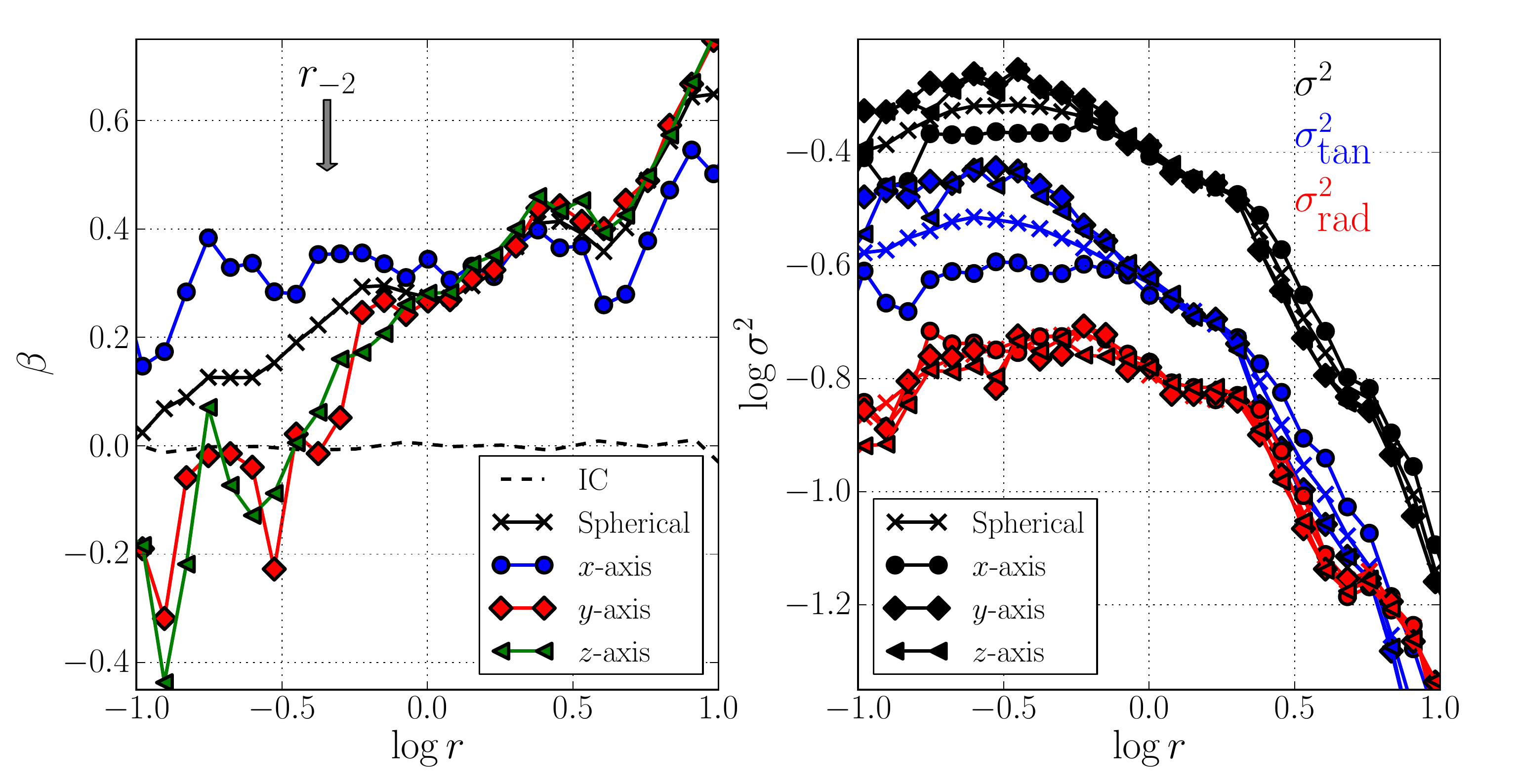}
\caption{Properties of particles in cones along the $x$-, $y$- and $z$-axis are analysed together with the particles in spherical bins. The merger has no impact parameter, and the collision axis is the $x$-axis. \new{$\beta_\text{initial}=0$} for the haloes. The \emph{left panel} shows the velocity anisotropy in each cone, and the \emph{right panel} shows the total, the tangential and the radial velocity dispersion in each cone. The grey arrow shows the radius, where $\gamma=-2$.}
\label{BetaAndSigma_1HqIso_Impact0}
\end{figure}

First the remnant of the major merger simulation for the \new{$\beta_\text{initial}=0$}-haloes without an impact parameter is studied. Figure~\ref{BetaAndSigma_1HqIso_Impact0} (\emph{left panel}) shows the velocity anisotropy in cones along each axis. Also shown is the spherically averaged value of $\beta(r)$. In the cone pointing in the direction of the collision axis ($x$-axis) the $\beta$-profile is clearly different from $\beta$ in the two perpendicular cones along the $y$- and $z$-axis. In the inner parts (with $r<1$) $\beta$ is almost constant ($\beta \simeq 0.2-0.4$) along the collision axis, whereas it is an increasing function of radius in the two other cones.

In the figure an arrow marks the radius of $r_{-2}$, which has been found by examining $\gamma(r)$. For a Hernquist profile, which was used as initial conditions in the simulation, $r_{-2}=r_\text{s}/2$. In cosmological simulations the virial radius is typically $10 r_{-2}$ \citep{2008MNRAS.391.1940M}, and in our simulations this is also roughly the radius, \new{to which structures are equilibrated.}

The radial ($\sigma_\text{rad}$), tangential ($\sigma_\text{tan}$) and total ($\sigma$) velocity dispersions for each of the three cones  are shown in the \emph{right panel} of Figure~\ref{BetaAndSigma_1HqIso_Impact0}. Most striking is the significantly lower value of the tangential dispersion along the collision axis. The radial dispersion profile is nearly the same in all the cones.

\begin{figure}
\centering
\includegraphics[width=\textwidth]{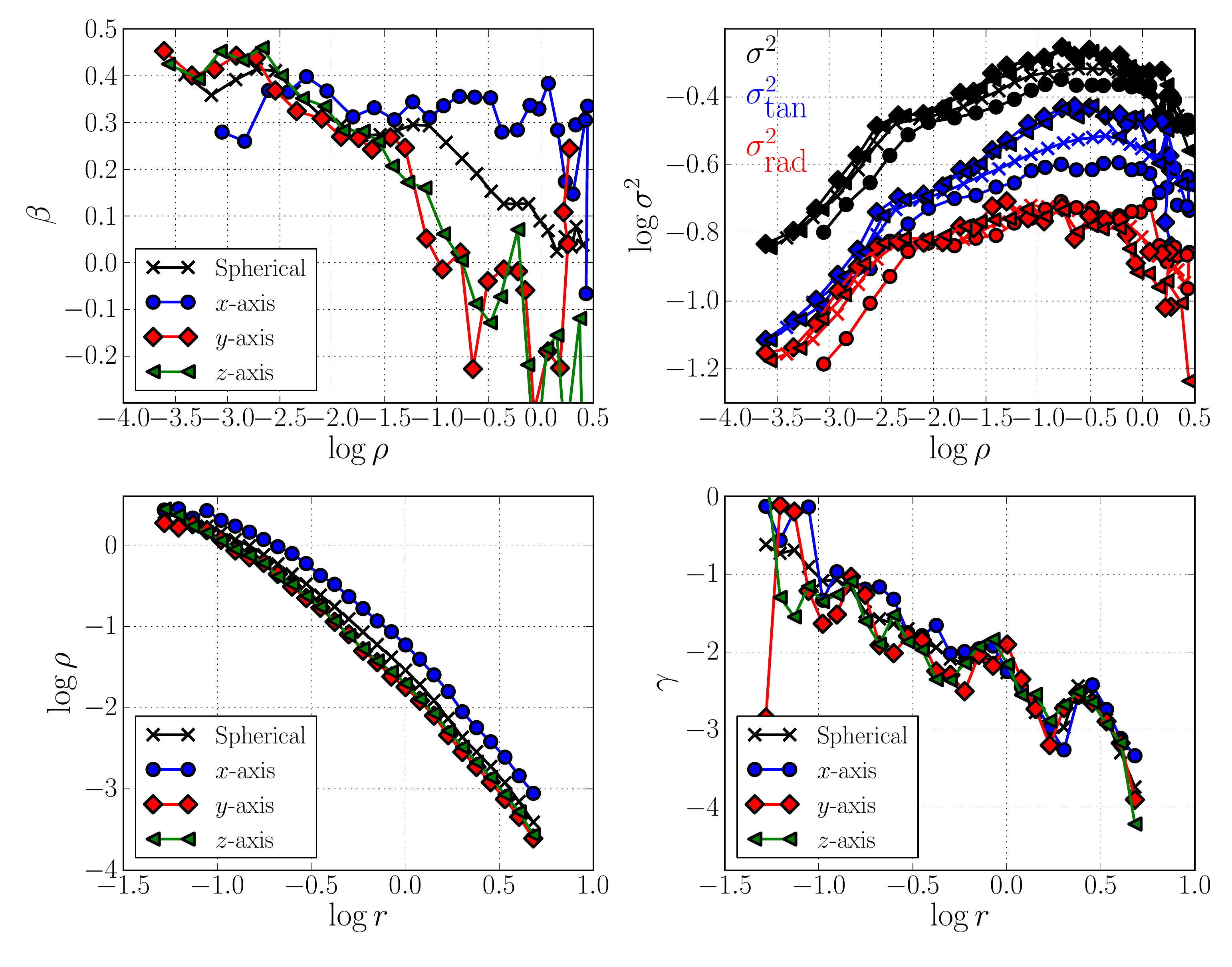}
\caption{$\beta(\rho)$, $\sigma^2 (\rho)$, $\rho(r)$ and $\gamma (r)$ in the different cones for a major merger without an impact parameter.}
\label{BetaAndSigma_Rho_1HqIso_Impact0}
\end{figure}

Figure~\ref{BetaAndSigma_Rho_1HqIso_Impact0} shows how the velocity anisotropy and the velocity dispersions (\emph{upper panels}) depend on the local density in each cone. Using $\rho$ as the abscissa corresponds to binning particles of equal density. $\beta(\rho)$ and $\sigma^2 (\rho)$ are clearly different in the three cones, so the $\beta$-profiles and the velocity dispersions are not constant along contours of equal density.

Another observation (\emph{lower panels}) is that the \new{slopes}, $\gamma(r)$, of the density profiles are similar along the different axes, but the actual value of $\rho$, is significantly larger in direction of the collision axis.

\subsubsection{\new{Comparison with the Via Lactea II halo}}

\new{In the Via Lactea II simulation \citep{2008Natur.454..735D} local properties of a Milky-Way-like halo have been examined \citep{2009MNRAS.394..641Z}. A direction-dependent $\beta$-profile is found in the inner parts (at 8 kpc) of the halo: along the major axis $\beta$ is larger ($0.2 \lesssim \beta \lesssim0.45$), than along the intermediate axis ($-0.1\lesssim\beta\lesssim 0.25$) and the minor axis ($-0.65 \lesssim\beta\lesssim -0.1$).}


\new{Our merger remnant is elliptical with a major axis pointing along the $x$-direction (see Section~\ref{shape} for details about the shapes). We can therefore confirm the finding (from \citep{2009MNRAS.394..641Z}) that $\beta$ is largest along the major axis.}

\subsubsection{The redistribution of \new{kinetic energy}}

Imagine two merging haloes, $\mathcal{A}$ and $\mathcal{B}$, approaching each other with a velocity similar to the escape velocity at the cluster's separation distance. Seen from the center of mass of halo $\mathcal{A}$, the particles from halo $\mathcal{B}$ have much \new{larger kinetic energies along the collision axis than along the perpendicular axes due to the relative motion of the haloes. The reason why merger remnants have different $\beta$-profiles in different cones is likely that merging processes are inefficient at transferring the kinetic energy along the collision axis, into kinetic energy along the other axes.} Many particles in a merger remnant are likely left on non-spherical orbits, where they oscillate back and forth along the collision axis. Such a scenario is perfectly consistent with what we see in Figure~\ref{BetaAndSigma_1HqIso_Impact0}.

\begin{figure}
\centering
\includegraphics[width=\textwidth]{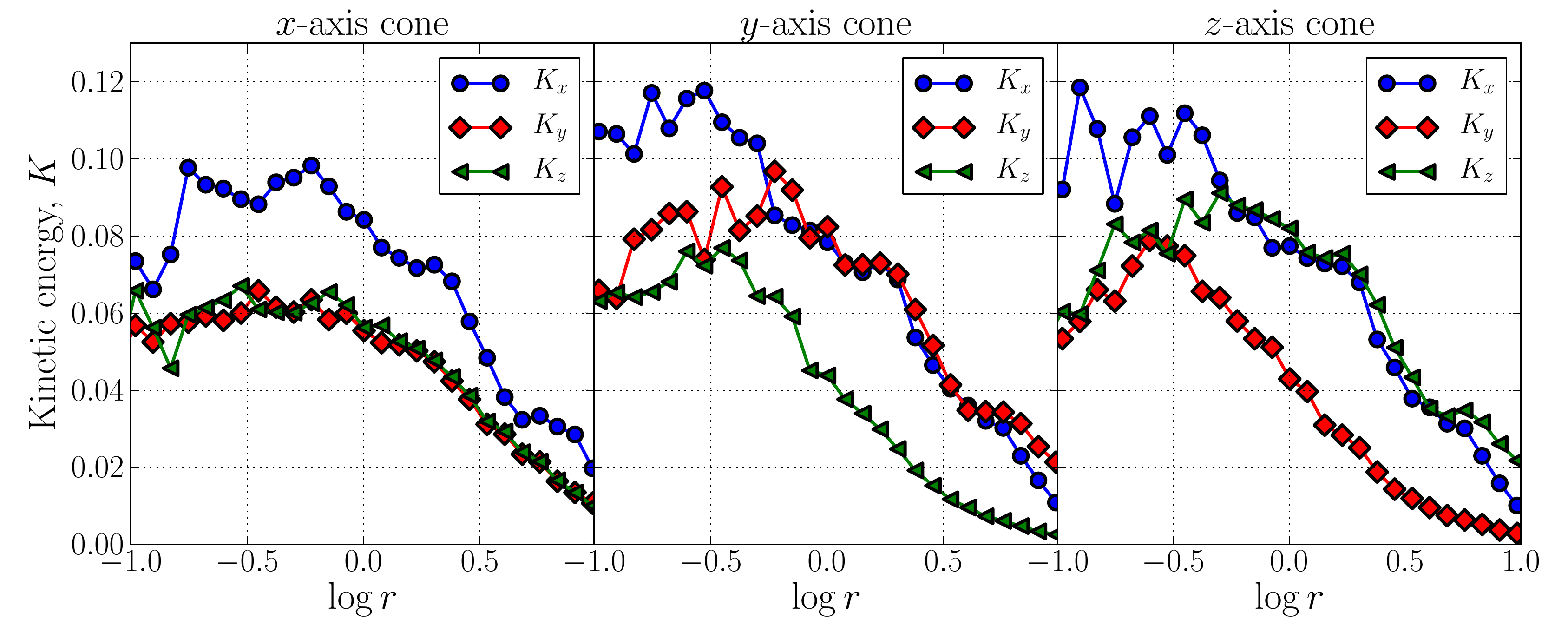}
\caption{\new{The mean kinetic energies in the $x$-, $y$- and $z$-direction along the three axes. $K_x$ in the cone along the $y$-axis is larger than $K_y$ in the cone along the $x$-axis. This shows that there is more kinetic energy along the collision axis of the merger remnant than along the other axes}.}
\label{Fig03B_K}
\end{figure}

\new{To test this explanation further the kinetic energies ($K_x$, $K_y$ and $K_z$) along each axis are calculated in each cone, see Figure~\ref{Fig03B_K}. Along the collision axis ($x$), we find that $K_y\simeq K_z$ at all radii. Along the $y$-axis, $K_x$ is larger than $K_z$ at all radii. This supports our claim, that the kinetic energy is not totally redistributed from the direction of the collision axis to the perpendicular directions in a merger.}

\new{An observation is that $K_x \simeq K_y $ in the outer parts of the cone along the $y$-axis. This means that in a cone perpendicular to the collision axis, we have that $\sigma (\text{collision axis}) \simeq \sigma_\text{rad}$ in the outer parts.}

\subsection{The effect of an impact parameter}

The presence of impact parameters breaks the rotation symmetry around the collision axis (which was present in the simulation without impact parameters), since the merging haloes are spiralling around each other in the $x$-$y$ plane.

\begin{figure}
\centering
\includegraphics[width=\textwidth]{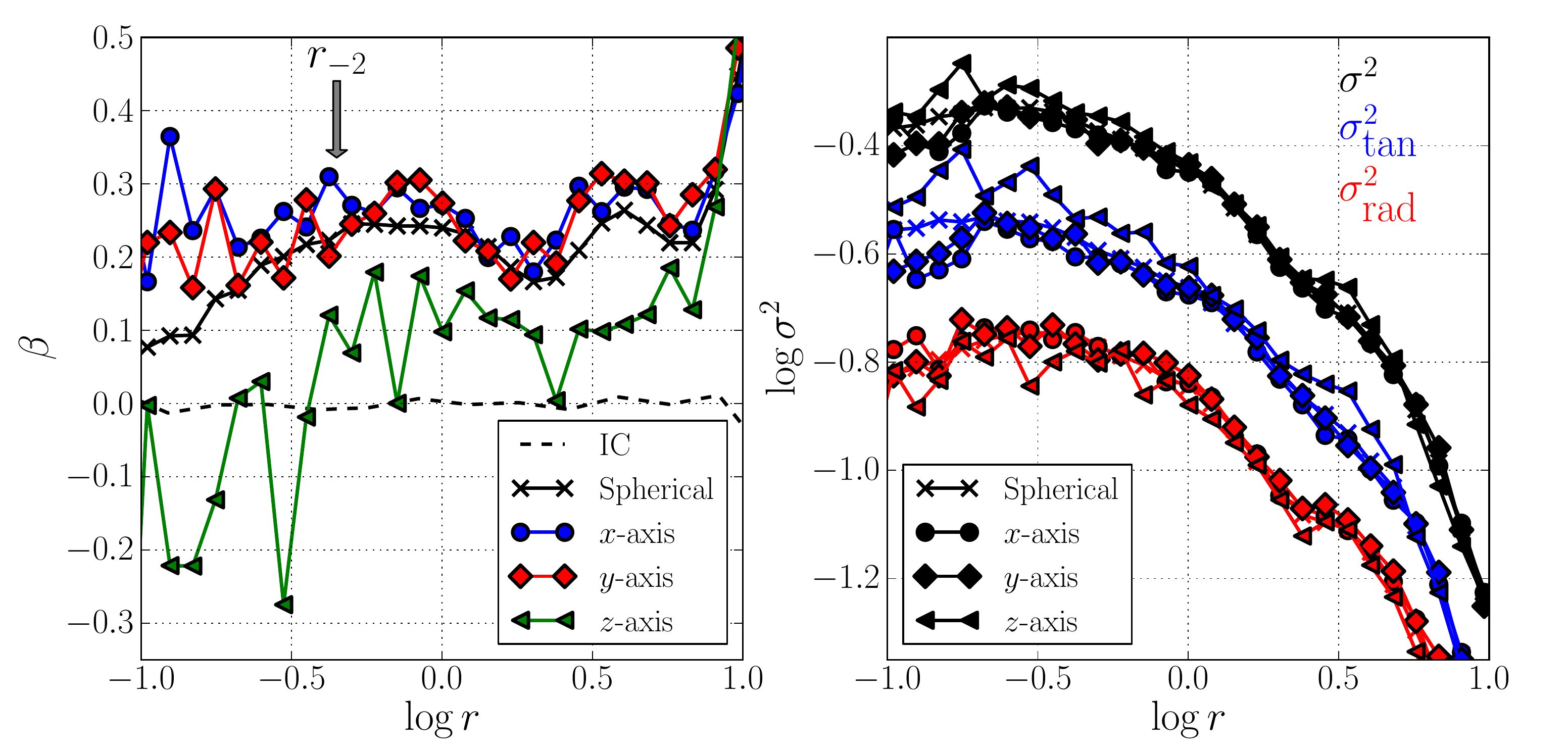}
\caption{The velocity anisotropy and the velocity dispersion for the major merger simulation with an impact parameter. The merging process is taking place in the $x$-$y$ plane.}
\label{BetaAndSigma_1HqIso_Impact10}
\end{figure}

Figure~\ref{BetaAndSigma_1HqIso_Impact10} shows the anisotropy and the velocity dispersions, when an impact parameter of $10r_\text{s}$ is present. The rotation velocity at a given radius was subtracted before the velocity dispersion in a given bin was calculated. The velocity anisotropy is constant \new{and positive} in the inner regions in both the $x$- and in the $y$-direction, and negative in the $z$-direction. It is different from the simulations without impact parameters, where $\beta$ was negative in both the $y$- and the $z$-direction. \new{The spherically averaged $\beta$-profile has a non-monotonic shape, similar to what is found in other studies of major merger remnants (e.g. \cite{2007MNRAS.376.1261M}).}

The velocity dispersions are similar in all the three cones, except for the inner parts, where the cone in the $z$-direction has a large\new{r} tangential and a lower radial dispersion.

\subsection{Major mergers with Osipkov-Merritt haloes}

\begin{figure}
\centering
\includegraphics[width=\textwidth]{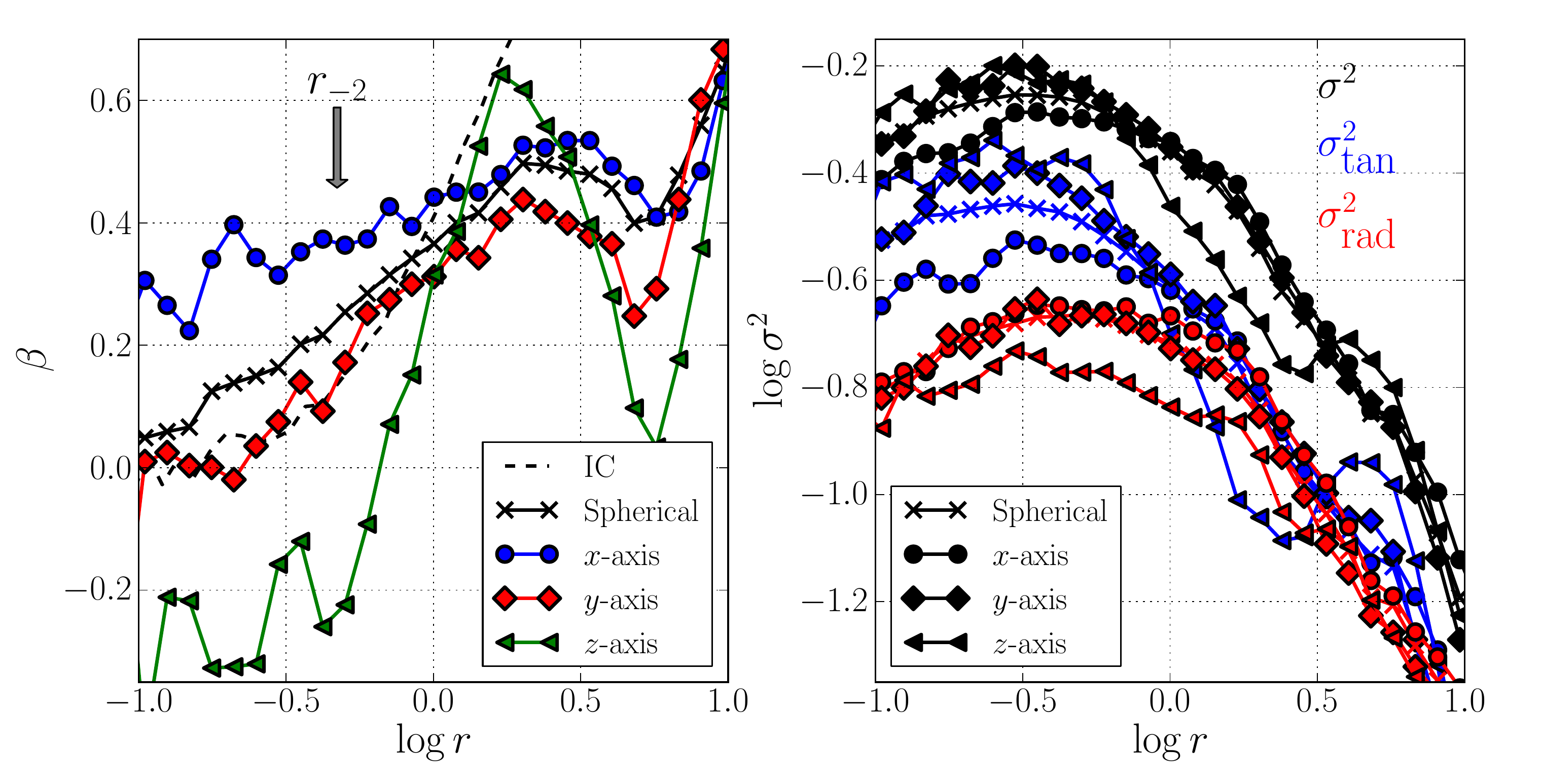}
\caption{Same as Figure~\ref{BetaAndSigma_1HqIso_Impact0}, but with initial haloes following Osipkov-Merritt models.}
\label{BetaAndSigma_1HQOM_Impact0}
\end{figure}

Figure~\ref{BetaAndSigma_1HQOM_Impact0} shows the remnant of two Osipkov-Merritt haloes collided without an impact parameter. In this case the merger remnant again had different velocity anisotropies and velocity dispersions in the different cones. 

Before the progenitor cores touched each other, the two haloes were clearly more elongated, when visually inspected, than for the $\beta_\text{initial}=0$ mergers. Instabilities, such as the radial orbit instability, are therefore likely play an important role, and such instabilities might be the reason why the velocity anisotropies are not the same in the $y$- and the $z$-direction, which would be expected from the symmetries of the initial conditions. 

\subsection{Minor mergers}

\begin{figure}
\centering
\includegraphics[width=\textwidth]{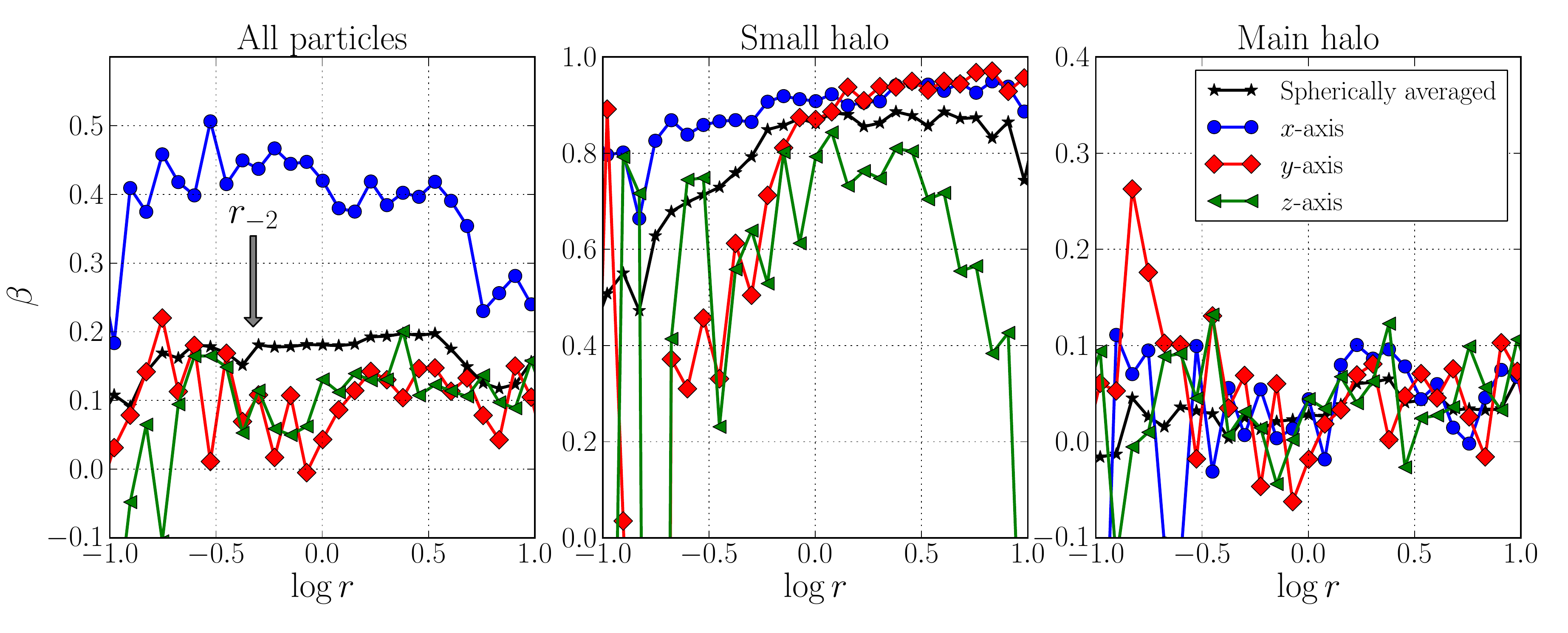}
\caption{Particles in a remnant of a 1:10 merger without an impact parameter are analysed in cones. The \emph{left panel} shows all the particles in the simulation, the \emph{central panel} shows particles that started in the small halo, and the \emph{right panel} shows particles from the main halo.}
\label{BetaSmall}
\end{figure}

Figure~\ref{BetaSmall} shows the velocity anisotropies in a 1:10 merger remnant for a simulations without an impact parameter. Three sets of particles are analysed: i) all the particles in the simulation (\emph{left panel}), ii) the particles that started in the small halo (\emph{central panel}) and iii) the particles originating from the main halo (\emph{right panel}). The particles from the main halo remain at $\beta=0$, and are almost unaffected by the merging process. The particles from the small halo are in orbits with $\beta =  0.3-1.0$. When all particles are analysed $\beta$ is positive ($\beta \simeq$0.4) along the collision axis, but a nearly isotropic velocity distribution ($\beta\simeq 0.1$) are present in the perpendicular directions.

\begin{figure}
\centering
\includegraphics[width=\textwidth]{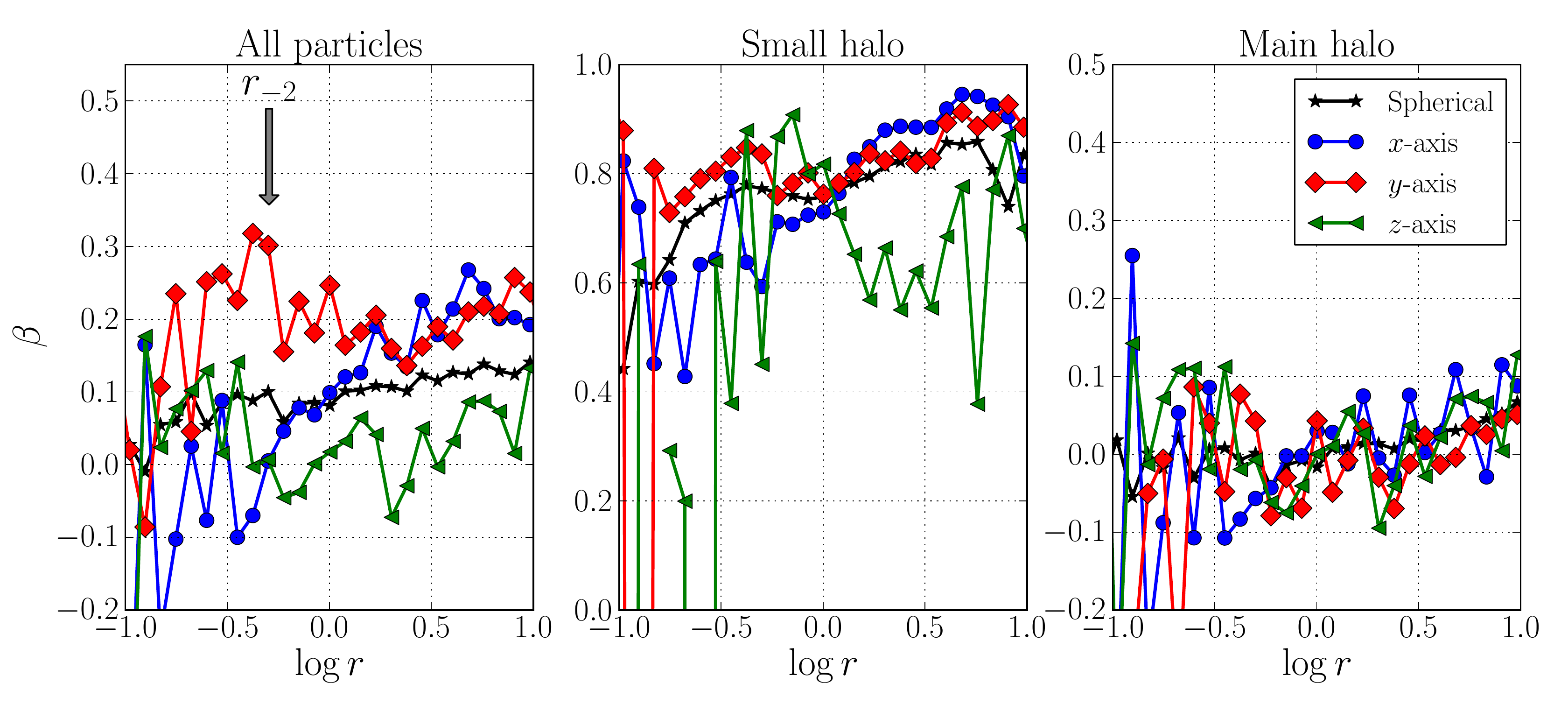}
\caption{Same as Figure~\ref{BetaSmall}, but with an impact parameter along the $y$-axis.}
\label{BetaSmallImpact}
\end{figure}

With an impact parameter present the velocity anisotropies are as shown in Figure~\ref{BetaSmallImpact}. The main halo particles remain at $\beta=0$, but the spherically averaged $\beta$-profiles are again positive due to the contribution from the particles from the small halo. Also note that the velocity anisotropy in the $y$-direction is larger than in the simulation without an impact parameter.

The complicated direction-dependence of the velocity distributions in the minor merger remnants, is likely similar to the direction-dependent velocity structure in the substructure of cosmological simulations \citep{2011arXiv1105.4166L}.

\section{The effect of an accretion process}\label{Accretion}

Like mergers, smooth accretion of matter is a very important process for the build up of structure in the universe. We now want to study how the asymmetric $\beta$-profiles in merger remnants are affected by a spherically symmetric perturbation that mimics smooth accretion.

In the remnant of the major merger simulation without an impact parameter (from Section~\ref{major_mergers}), we inserted particles in a spherical shell in the region, $4.8<r/r_{-2}<15$, with a density profile following, $\rho (r) \propto r^{-2}$. Each particle has a radial infall velocity identical to the escape velocity at \new{its position}. The number of particles in the shell and the mass of each particle are the same as in the major merger simulations ($2\times 10^6$ particles with a total mass of 2). The simulations were run for 300 time units, which corresponds to 250 dynamical times, $\sqrt{r^3/GM(r)}$, at $r_{-2}$ of the final structure, and 20 dynamical times at $10r_{-2}$.

\begin{figure}
\centering
\includegraphics[width=0.8\textwidth]{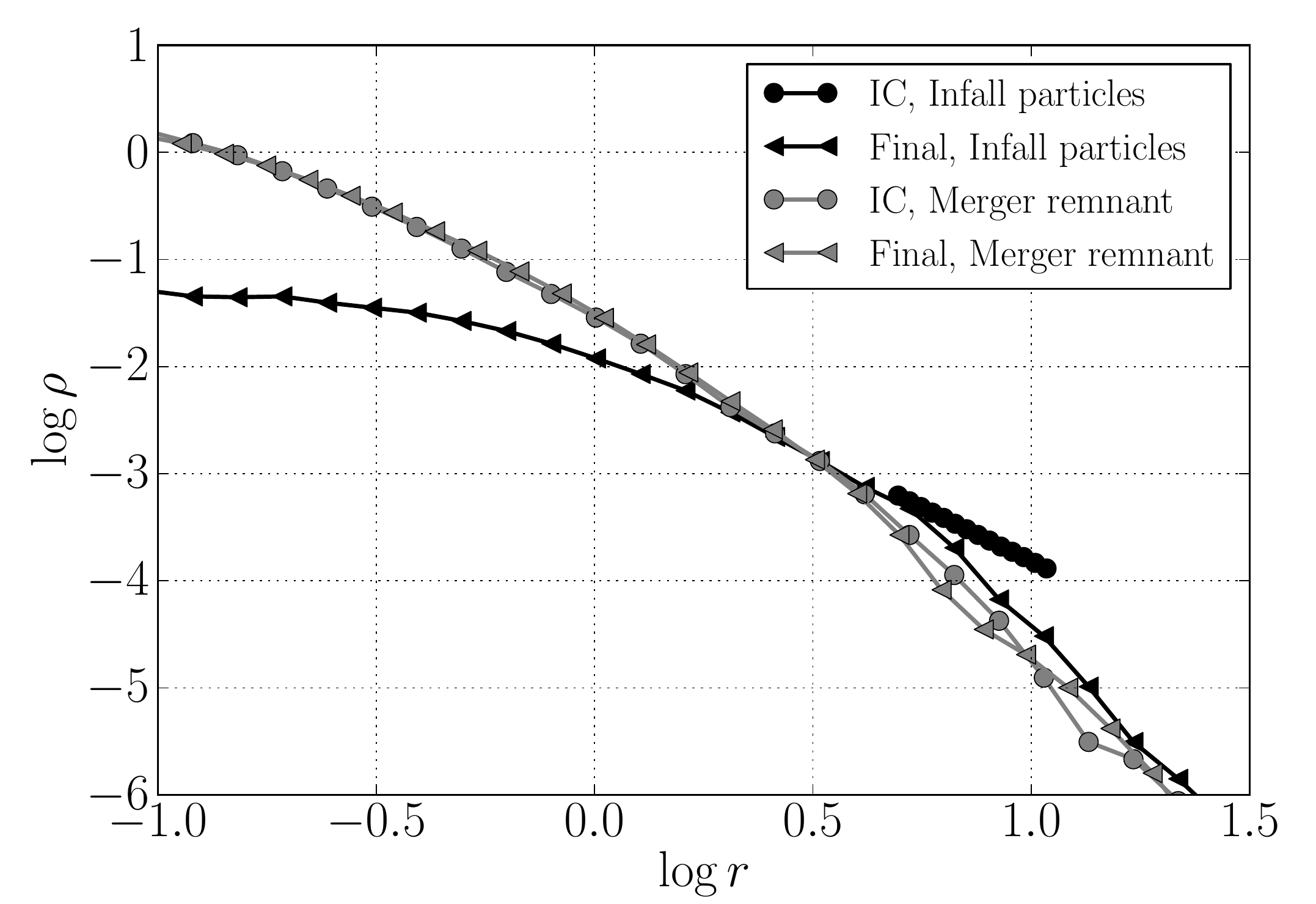}
\caption{The initial conditions and the results of the simulations, where a merger remnant (grey points) is perturbed by a process that mimics smooth accretion. A group of particles accretes onto the structure from a shell at radii $4.8<r/r_{-2}<15$ (black circles).}
\label{Rho_accretion}
\end{figure}

Figure~\ref{Rho_accretion} shows the evolution of the density profiles for the particles that started in the merger remnant, and for the infall particles that started in the shell. The infall particles only gives a minor contribution to the central densities of the final structure ($\lesssim 5\%$).

\begin{figure}
\centering
\includegraphics[width=\textwidth]{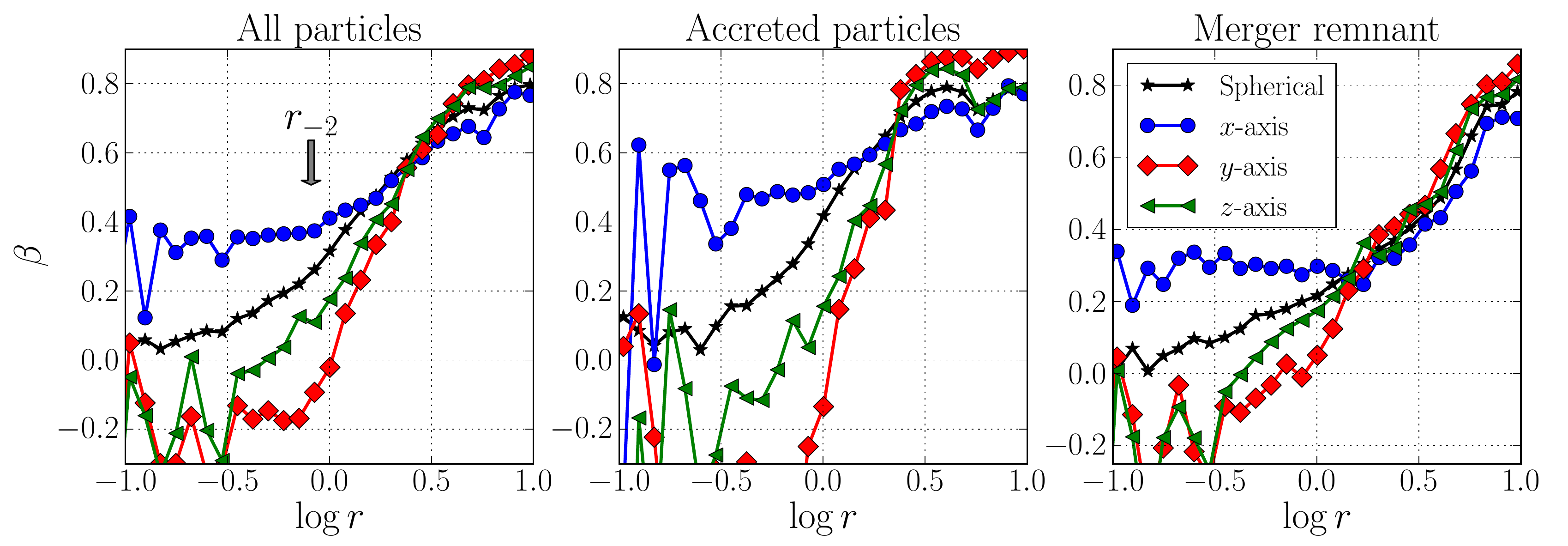}
\caption{The velocity anisotropy of the particles in the accretion simulation. The \emph{left panel} shows $\beta(r)$ for all the particles in the simulation, the \emph{central panel} shows the particles from the accretion shell, and the \emph{right panel} shows the particles that started in the merger remnant.}
\label{Beta_Accretion}
\end{figure}

Figure~\ref{Beta_Accretion} shows the anisotropy for the final structure for all the particles, the accreted particles and the particles that started in the merger remnant. For all sets of particles, $\beta$ is \new{largest} in the $x$-direction, and smaller in the directions perpendicular to the merger axis. \new{An interesting observation is that the $\beta$-profiles of the accreted particles are very similar to the $\beta$-profiles of the particles from the merger remnant.}

We conclude that the asymmetric velocity anisotropies remain asymmetric, when perturbed by this accretion process.

\section{Skymaps of the anisotropies} \label{skymap}

\new{So far we have focused on analysing particles in cones in the $x$-, $y$- and $z$-direction. We will now take a more detailed look on the angular dependence of the $\beta$-profiles. To do so we distributed 192 points on a sphere using the HEALPIX framework \citep{2005ApJ...622..759G}, and defined a cone pointing in the direction of each point. The apex angle was still 45$^\circ$, so the cone angle is larger than the size of one pixel.}

\begin{figure}
\centering
\includegraphics[width=\textwidth]{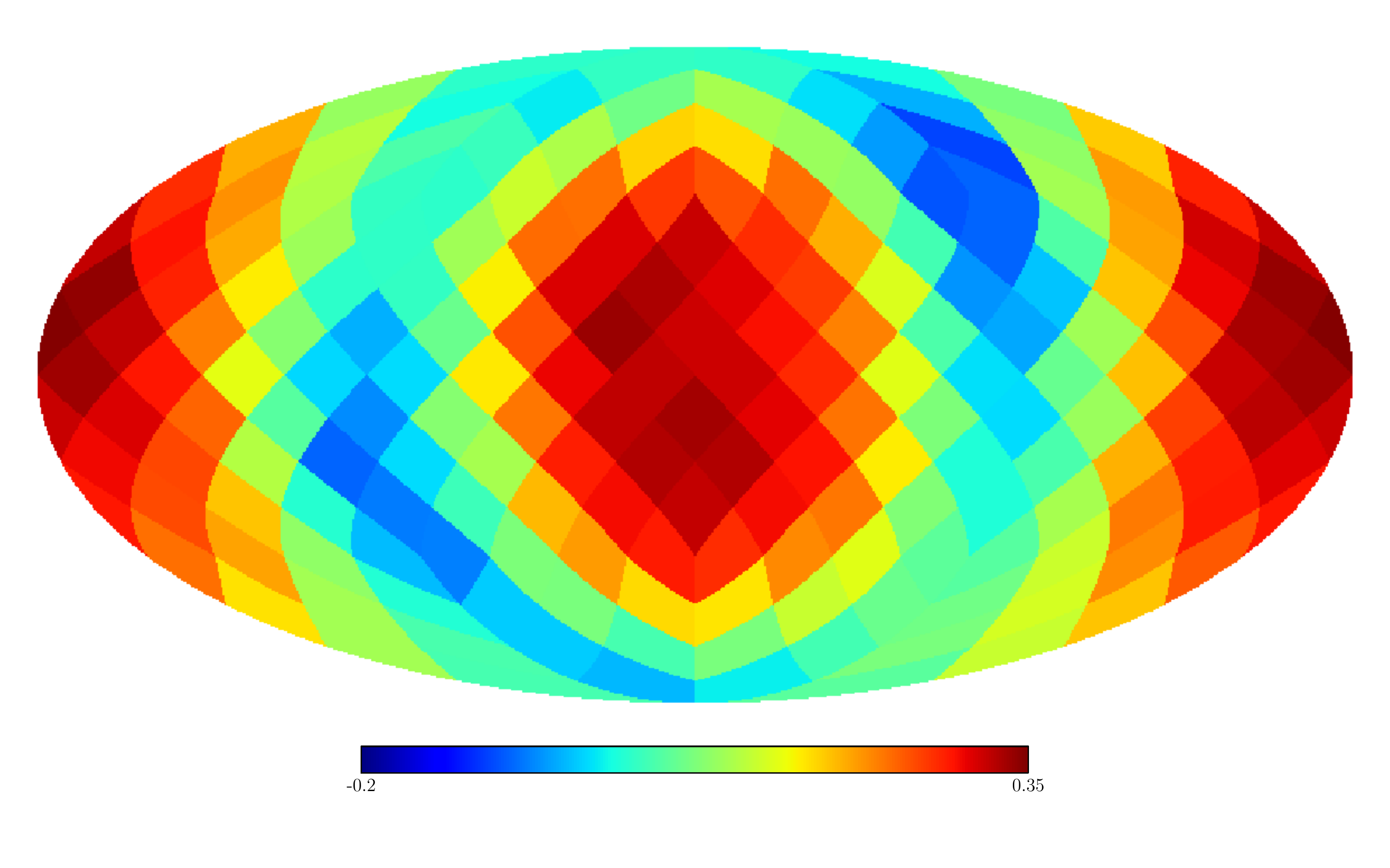}
\caption{Skymap of $\beta$ of the major merger with $\beta_\text{initial}=0$.}
\label{CMB1}
\end{figure}

\begin{figure}
\centering
\includegraphics[width=\textwidth]{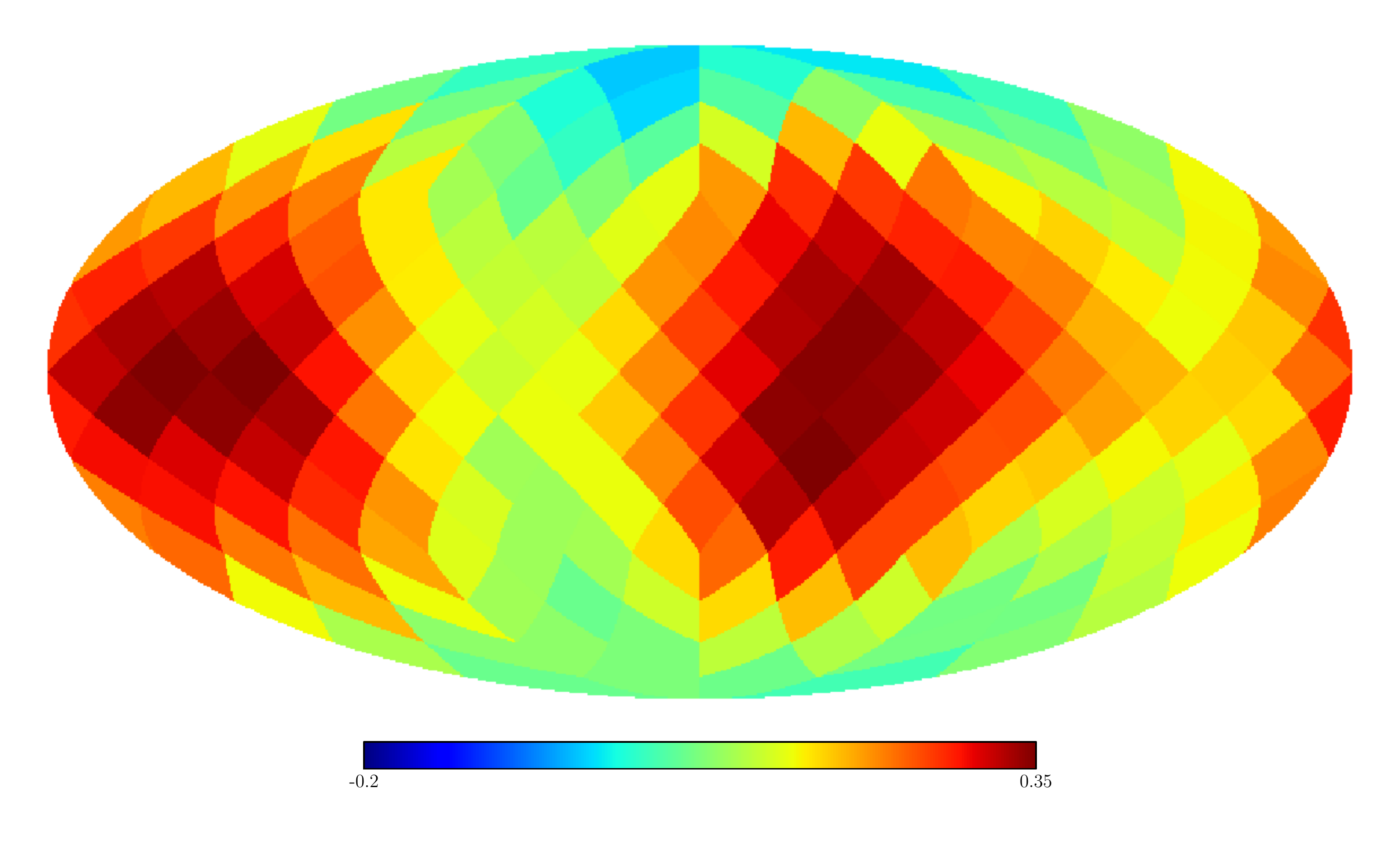}
\caption{Skymap of $\beta$ of the major merger with $\beta_\text{initial}=0$ and an impact parameter along the $y$-axis.}
\label{CMB2}
\end{figure}

\begin{figure}
\centering
\includegraphics[width=\textwidth]{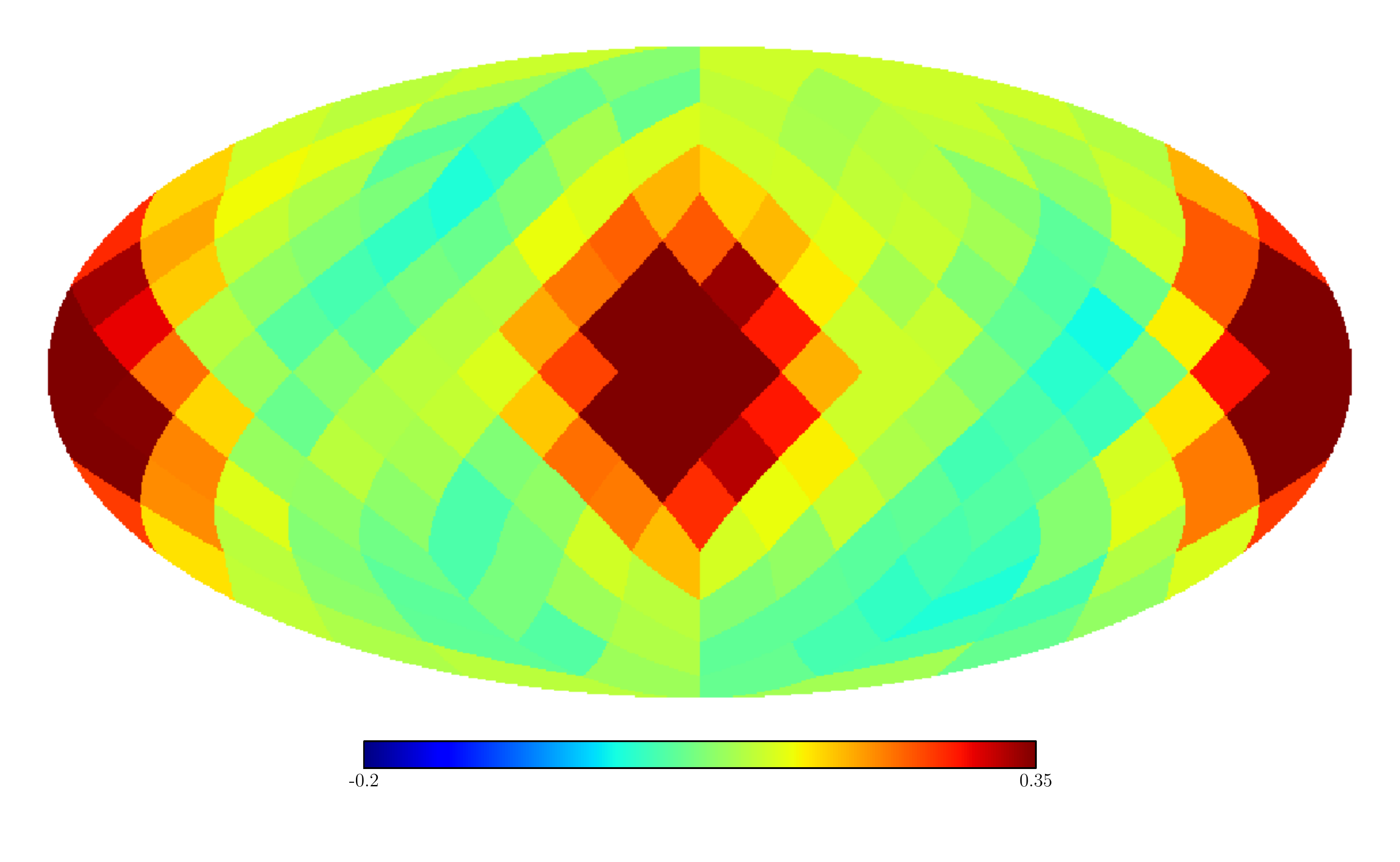}
\caption{Skymap of $\beta$ of the minor merger with $\beta_\text{initial}=0$ and no impact parameter.}
\label{CMB4}
\end{figure}

\new{Figure~\ref{CMB1} shows a Mollweide projection (which is heavily used in CMB analysis) of $\beta$ at a radius of $0.35$ for the major mergers with $\beta_\text{initial}=0$. Figure~\ref{CMB2} shows the same plot for the major merger with an impact parameter along the $y$-axis, and Figure~\ref{CMB4} shows the minor merger without an impact parameter.}

\new{In the Mollweide projection the positive $x$-direction is in the center of the plot, and negative $x$-direction are the points most to the right and left. The positive $z$-direction is the top point, and the negative $z$-direction is the bottom point. The $y$-axes are in the two points between the positive and negative $x$-axis (the positive $y$-axis is to the right, the negative to the left).}

\new{A visible effect is that the presence of an impact parameter breaks the symmetry along the $y$-$z$ plane in Figure~\ref{CMB2}. It is also clear that the minor merger, mostly affects $\beta$ along the infall direction (Figure~\ref{CMB4}).}

\section{$\beta$-$\gamma$ relations}\label{b-g-relations}

\begin{figure}
\centering
\includegraphics[width=\textwidth]{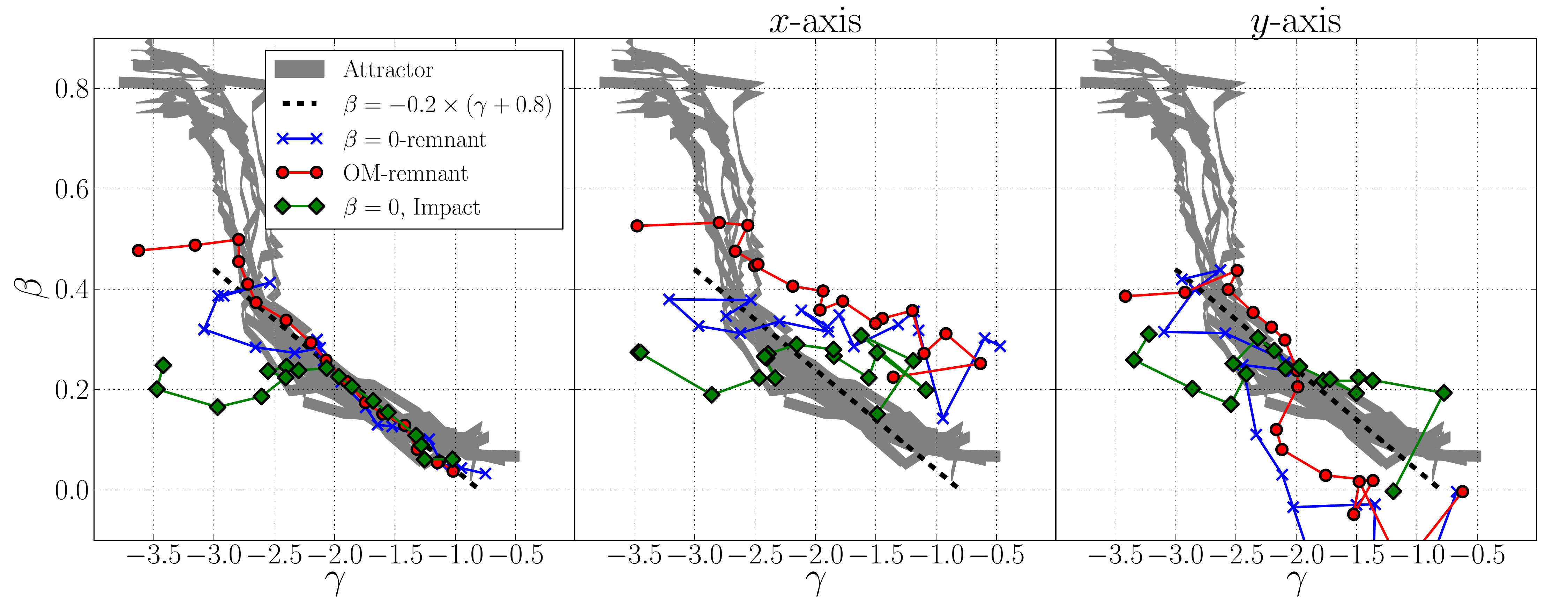}
\caption{$\beta (\gamma)$ profiles for the remnant of the major merger between the \new{$\beta_\text{initial}=0$} haloes (with and without an impact parameter), and the major merger with two Osipkov-Merritt haloes. The remnants are compared between two proposed $\beta(\gamma)$-relations: $\beta = -0.2\times (\gamma+0.8)$ \citep{2006JCAP...05..014H} and the \emph{attractor} (the grey lines are taken from \citep{2010ApJ...718L..68H}). The \emph{left panel} shows spherically averaged profiles, the \emph{central panel} shows a cone along the collision axis, and \emph{right panel} shows a cone perpendicular to the collision axis.}
\label{Fig10_GammaBeta}
\end{figure}

We will now compare the merger remnants with the $\beta$-$\gamma$ relation \citep{2006JCAP...05..014H}, $\beta = -0.2\times (\gamma+0.8)$, and the attractor \citep{2010ApJ...718L..68H,2012arXiv1204.2764B}. Figure~\ref{Fig10_GammaBeta} (\emph{left panel}) shows the spherically averaged $\beta (\gamma)$ profiles for the major merger remnants. The \emph{central panel} and the \emph{right panel} show the same for the cones along the $x$- and $y$-axis, respectively.

The $\beta$-$\gamma$ relations are clearly not obeyed in the two plotted cones, but the spherically averaged $\beta$-profiles are in good agreement with the two predictions in the inner parts with $\gamma>-2.2$. The outer parts with $\gamma<-2.2$ deviate from the relations. \new{It is seen that the spherically averaged $\beta$- and $\gamma$-profiles obey the $\beta$-$\gamma$-relations at the radii, where the $\beta$-profile is direction-dependent.}

We see that the spherically averaged $\beta(\gamma)$ profile in the outer parts is strongly dependent on a remnants merging history. This finding can explain the large scatter of $\beta(\gamma)$ between haloes in cosmological simulations \citep{2011MNRAS.tmp..937L}. In Figure~\ref{BetaAndSigma_1HqIso_Impact0}, \ref{BetaAndSigma_Rho_1HqIso_Impact0} and \ref{BetaAndSigma_1HQOM_Impact0}, we also find that $\beta(r)$ depends on the detailed merging history, so we conclude that the differences in $\beta(r)$ from halo to halo in cosmological simulations can be caused by their different merging histories.

\section{Pseudo-phase-space density profiles} \label{SectionPPSD}

\begin{figure}
\centering
\includegraphics[width=\textwidth]{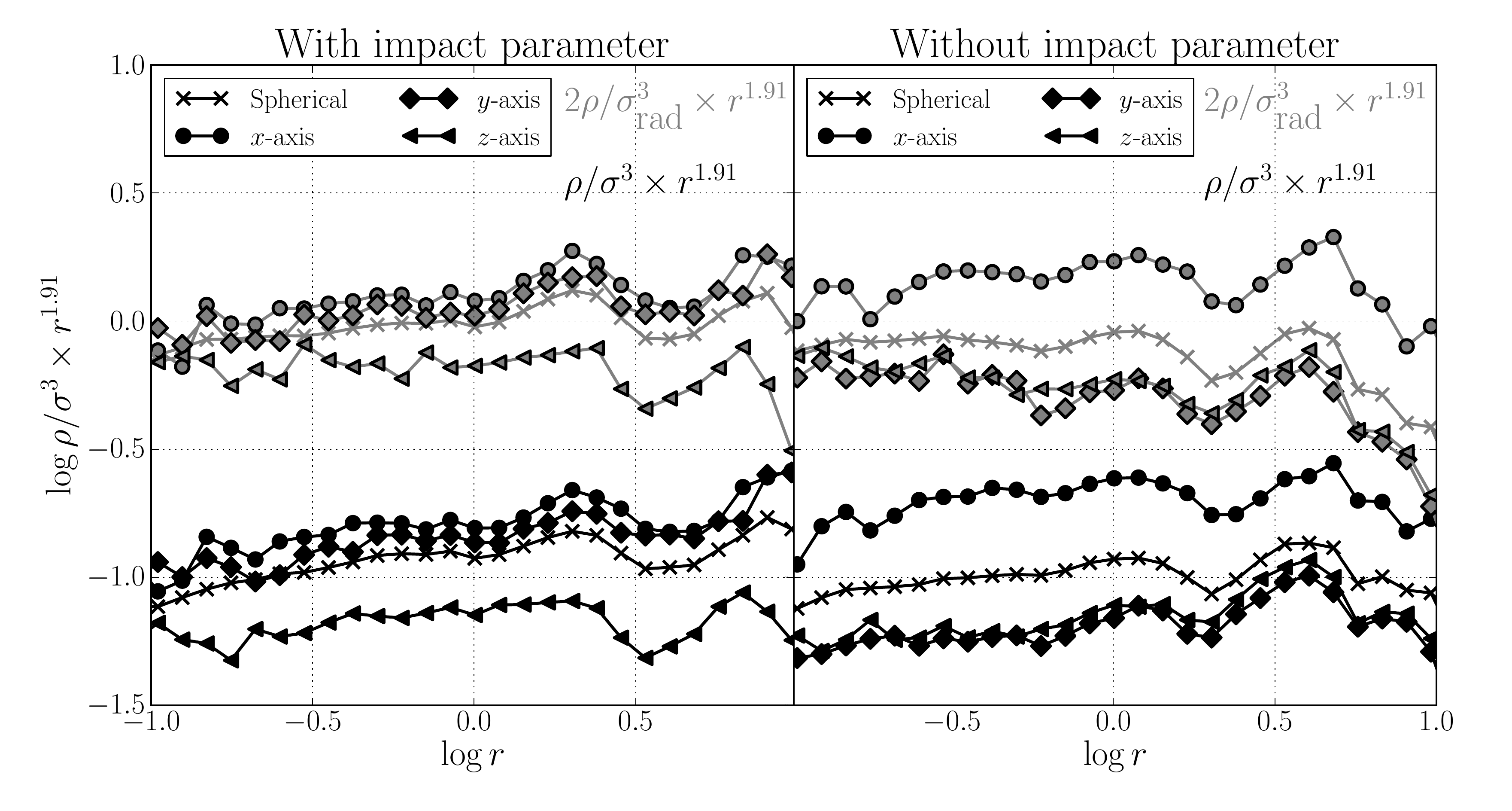}
\caption{The pseudo-phase-space density profiles, $\rho/\sigma_\text{rad}^3$ and $\rho/\sigma^3$, for the major merger remnant with an impact parameter (\emph{left panel}) and without and impact parameter (\emph{right panel}). $\rho/\sigma_\text{rad}^3$ is multiplied with a factor of 2 for presentation reasons.}
\label{PPSD}
\end{figure}

The pseudo-phase-space density profiles (PPSD's), $\rho / \sigma_\text{rad}^3$ and $\rho / \sigma^3$, of the major merger remnants with \new{$\beta_\text{initial}=0$} are shown in Figure~\ref{PPSD}. The profiles are scaled with a factor of $r^{1.91}$, which roughly would give a constant in cosmological simulations \citep{2011MNRAS.tmp..937L}. In both remnants the normalizations of the PPSD-profiles depend on the cone-direction. The best-fitting value for $\alpha$ only has a small variation from cone to cone.

In several studies dynamical constraints of dark matter haloes have been derived by assuming a radial power law behaviour of $\rho / \sigma_\text{rad}^3$ or $\rho / \sigma^3$ \citep{2004MNRAS.352L..41H,2005ApJ...634..756A,2005MNRAS.363.1057D}. In the merger remnants this assumption is however not correct, due to the different normalizations from cone to cone, so such an approximation does not describe the full dynamics of the remnants.

\section{The shape of the haloes}\label{shape}

\new{Is is established that a merger with two spherical haloes results in a triaxial halo with a major axis along the collision axis \citep{2004MNRAS.354..522M}. It has also been shown that the potential and velocity contours of a halo are more spherical than the density contours \citep{2006MNRAS.367.1781A, 2007ApJ...671.1135K}.}

\new{In our merger remnants we measure the minor to major axis ratio, $c/a$, by comparing density profiles in cones through points distributed uniformly of a sphere (with the same method as in Section~\ref{skymap}). To be explicit we will find the maximum and minimum values of the radius for a given density, and then calculate the ratio between these two radii. The shape will only be determined at $r=1$.}

\new{For all the mergers without an impact parameter the halo shapes were aligned with the collision axes. For the major mergers with $\beta_\text{initial}=0$ the axis ratio was 0.59, for the minor merger we found a ratio of 0.84, and for the Osipkov-Merritt model we found 0.53. For the major merger with an impact parameter we found a ratio of 0.7, and for the minor merger with an impact parameter we found 0.86.} 

\new{Our simulations are therefore consistent with the discovery in \citep{2009MNRAS.394..641Z}, in which it was discovered that $\beta$ is aligned with the shape of cosmological haloes, with positive $\beta$-values along the major axis, and negative $\beta$-values along the minor axis.}

\section{Conclusion}

We have analysed the velocity distributions in different cones centered on merger remnants. The velocity anisotropy profiles and the velocity dispersions in the different cones behaved differently. This was both the case for major mergers with and without impact parameters, and for minor mergers. We also demonstrated that these asymmetries are not washed out by a process that mimics smooth accretion. \new{Since mergers are frequent in the real universe, it is therefore not surprising that similar asymmetries are present in cosmological haloes \citep{2009MNRAS.394..641Z}.}

In the various merger simulations, the behaviour of $\beta(r)$ in the outer parts had a huge dependency on the initial conditions. We therefore conclude, that the different behaviour of $\beta(r)$ from halo to halo in cosmological simulations can be caused by their different merger histories.

Several studies \citep{1967MNRAS.136..101L,2000ApJ...531..739N,2009ApJ...694.1250H,2010ApJ...722..851H,2010ApJ...722..856W, 2010ApJ...725..282W,2011A&A...526A.147K,2012arXiv1206.5306H} have attempted to derive or characterise distribution functions of completely relaxed haloes from first principles. We have demonstrated that merger remnants have the merger history encoded in their velocity anisotropy profiles, and they are \new{therefore not expected to follow simple distribution functions, where $\beta$ is constant within spherical bins or along the isodensity contours}. Furthermore it is also clear that the $\beta$-$\gamma$ relation or the \emph{attractor} are not obeyed in all cones in merger remnants. We do, however, find that the spherically averaged properties of haloes obey these relations in the inner parts.

Finally, we note that the merger history likely is important for observational aspects of the dark matter haloes. This is exemplified through a correlation between the asymmetric nature of the velocity anisotropy of haloes and the surrounding large-scale structure \citep{Wojtak}, along which matter accretion typically occurs \citep{2011MNRAS.411.1525L}, and the differences in the line-of-sight velocity dispersions of galaxies along the major and minor axes of galaxy clusters \citep{Skielboe}.

\acknowledgments

We thank A. D. Ludlow, J. Samsing, A. Skielboe and R. Wojtak for useful discussions. The Dark Cosmology Centre is funded by the Danish National Research Foundation. The simulations were performed on the facilities provided by the Danish Center for Scientific Computing.

\def\aj{AJ}
\def\araa{ARA\&A}
\def\apj{ApJ}
\def\apjl{ApJ}
\def\apjs{ApJS}
\def\apss{Ap\&SS}
\def\aap{A\&A}
\def\aapr{A\&A~Rev.}
\def\aaps{A\&AS}
\def\mnras{MNRAS}
\def\na{New Astronomy}
\def\jcap{JCAP}
\def\nat{Nature}
\def\pasp{PASP}
\def\aplett{Astrophys.~Lett.}
\def\physrep{Physical Reviews}
\bibliographystyle{JHEP}

\providecommand{\href}[2]{#2}\begingroup\raggedright\endgroup

\end{document}